\documentclass[CMYK,NoSecthm]{jcomsec_author}  
\usepackage{savesym}
\savesymbol{AND}
\usepackage[linkcolor=black,citecolor=black,colorlinks=false,bookmarksnumbered]{hyperref} 
\usepackage{multicol}
\usepackage{enumerate}
\usepackage{amssymb}
\usepackage{multirow}
\usepackage{amsmath}
\usepackage{graphicx}
\usepackage{subcaption}
\usepackage{setspace}
\usepackage{mathtools}
\usepackage{colortbl}
\usepackage{color}
\usepackage{ctable}
\usepackage{wrapfig}
\usepackage[numbers,sort&compress]{natbib}
\usepackage{hypernat}
\usepackage[all]{hypcap}
\usepackage{array}
\usepackage{microtype}
\usepackage{doi}

\DisableLigatures[f]{encoding = *, family = * }

\newcolumntype{X}[1]{>{\centering\arraybackslash}m{#1}}

\usepackage[autostyle]{csquotes}

\begin{document}
\begin{frontmatter}
%
\title{DiCuPIT: Distributed Cuckoo Filter-based Pending Interest Table }

\author[DLS]{Arman Mahmoudi}
\ead{Arman.Mahmoudi@outlook.com \rm(A. Mahmoudi)}
\author[DLS,CorAuth]{Mahmood Ahmadi}
\ead{m.ahmadi@razi.ac.ir \rm(M. Ahmadi)}
\address[DLS]{Department of Computer Engineering and Information Technology, Razi University, Iran.}
\corauth[CorAuth]{Mahmood Ahmadi.}

%


\begin{abstract}
Named data networking is one of the recommended {\color{red}architectures} for the future of the Internet. In this communication architecture, the content name is used instead of the IP address. To achieve this purpose, a new data structure is added to the nodes of named data networking which is called Pending Interest Table (PIT). Scalability, memory consumption, and integration are the significant challenges in PIT design {\color{red} as} it needs to be updated for each packet, and it saves the name of the packet. This paper introduces a new data structure for PIT called DiCuPIT. DiCuPIT is a distributed data structure for the PIT table, {\color{red} that works} based on the Cuckoo filter and can cover the three  features as above-mentioned. {\color{red} By} implementing this PIT, {\color{red} the lookup} time shows {\color{red} a 36\% reduction} compared to the methods based on the Bloom filter and 40\% based on hash tables. Moreover, the memory consumption is reduced by 68\% compared to the hash tables-based mechanisms and 31\% compared to the methods based on the Bloom filter.
\end{abstract}

\end{frontmatter} 

\section{Introduction}

 NDN is one of the recently proposed architectures for the future Internet that uses the content name rather than the host address in the network. All communications in these networks are based on two types of packets: Data and Interest \cite{ref1}. Both packets carry a name and are transmitted over the network based on these names \cite{ref1}. The consumer sends the content name to the network into an interest packet. Then, each node forwards in reverse to the consumer in the network that contains the data packet corresponding to the interest packet \cite{ref1}. NDN has a new table for maintaining the interest packet, which the data packets corresponding to them {\color{red} have} not still arrived at the node of the NDN network. This table is known as {\color{red} the PIT} \cite{ref1}, \cite{ref2}. When the interest packets are received, the PIT table makes an entry or updates it. In return, {\color{red} upon receipt of a data packet}, the node of the NDN network must find the interest packet corresponding to it in the PIT table, and {\color{red} forward} the data packet to the incoming interfaces of {\color{red} the received }interest packet, and finally {\color{red} remove} the intended entry in {\color{red} the PIT} table. Thus, the PIT table is strongly dynamic \cite{ref5}. Each entry in {\color{red}the PIT} table is stored as {\color{red} follows}; (content-name, list-faces, list-nonces, expiration) \cite{ref7}, \cite{ref3}. Searching is done by content-name in NDN.
  {\color{red}In the NDN,} the content-name is the same as the URL address, and it does not have a constant length like IP-based networks \cite{ref1}, so it is more complex and requires more memory. 
{\color{red} On the other hand, with the PIT implementation, as more packets of  interest  are stored in the PIT table over time, {\color{cyan} less time is spent for searching, which reduces the response time in the network}, thereby reducing response time across the  network.} Therefore, using data structures should not require much memory, and it reduces searching time and overall response time in return. The required memory for the PIT table is $O(n^7)$ \cite{ref4}, \cite{ref5}. Effective implementations {\color{red} on this topic have been made in} \cite{ref-new9} which  uses the hashing technique to solve the PIT problem. Although {\color{red}they have achieved} effective results, they have resolved some of the PIT needs. In \cite{ref5}, the authors {\color{cyan} have attempted} to design a distributed PIT table in which each interface {\color{red} in} the network's node contains a sub-table based on the Bloom filter. Despite {\color{cyan} the good PIT results with} this method, the problem is that {\color{red} the lookup should be performed on entire} sub-tables, which requires more response time. Generally, the PIT table has not been considered in this method. For instance, when an interest packet enters the PIT table, if exactly another packet with the same name enters from another face, these packets will be stored in different sub-tables, and information redundancy will happen. To cope with this problem in the PIT table, we propose a PIT table based on the Cuckoo filter, which is called DiCuPIT. In {\color{red} the DiCuPIT} designing, PIT is divided into $k$ sub-tables {\color{red} where} $k$ is the number of NDN router interfaces. In fact, this data structure is a distributed PIT which allocates a sub-table named $DiCuPIT_i$ to each interface. Each one of these $DiCuPIT_i$s has Cuckoo filter properties. These properties help us to be able to integrate a PIT while having a low lookup time and less memory consumption than similar ones. Integration means {\color{red} that the information of the packets of interest is stored} in such a way that other faces avoid storing the same packet in DiCuPIT to get the integration purpose in the PIT table. In addition to $DiCuPIT_i$s, we have {\color{red} considered another table called GlobalCu for each sub-table} to reduce lookup time named GlobalCu. According to Cuckoo filter properties and having GlobalCu, integration of PIT is guaranteed with this method. The results {\color{red} of the} experiments show that this method can reduce memory consumption by 31 \% compared to DiPIT and by 68\% compared to the hash tables-based methods, while {\color{red} the lookup time is reduced} compared to other methods. The novelty of this paper is explained as follows:
 \begin{itemize}
 \item Proposal of {\color{red} the distributed} DiCuPIT data structure based on the Cuckoo filters.
 \item Utilization of the DiCuPPIT data structure to improve the performance of the PIT table of NDN networks.
 \item Evaluation of the proposed DiCuPIT data structure in a typical NDN network. 
 \end{itemize}
 
 The following sections of this paper were organized as follows: Section \ref{relatedworks} reviews recent works on PIT table design. Section \ref{NDN} presents a brief concept of name data networking and probabilistic filters. Section \ref{dicupit} describes the proposed distributed Cuckoo filter-based PIT. Section \ref{results} presents the evaluation results. Section \ref{conclusion} concludes the paper. 

\section{Related Works}
\label{relatedworks}

This section, presents a brief review of the most significant works in PIT tables.

In \cite{ref-new1}, PIT control management (PITCM) to mitigate PIT overflow and {\color{red} improve} the PIT utilization is proposed. The PITCM utilizes a thresholding mechanism in interest lifetime in an intelligent manner, a policy {\color{red} for determining the highest} lifetime interest, and an adaptive virtualized PIT. The prediction of PIT conditions e.g., early PIT overflow prediction and reaction, is the responsibility of the adaptive virtual PIT block.
The smart thresholding of interest lifetime adjusts the lifetime value of received interest packets.
 Management of PIT entries in an efficient way is controlled by highest lifetime latest the request policy. The flow PIT sizing problem of the forwarding systems in the NDN networks {\color{cyan}is addressed} by this work.

In \cite{ref-new2}, to increase PIT efficiency by reducing the residence time of non-responded the PIT entries, round trip time-aware pending interest PIT (RAPIT) is proposed. The round tripe time-aware pending interest table (RAPIT) presents an approach to increase the performance of PIT by decreasing the residence time of non-responded PIT entries. The authors used a technique to determine the residence time of PIT entries based on the estimation of the return time of {\color{cyan} the data packets}.
 To measure the return time of the data packet, they mark the initial request and the corresponding data packet, and then store the measured time in the forwarding information base (FIB) of the data packet for future use. In addition, they use a PIT replacement strategy when the PIT is completely filled.

In \cite{ref5}, the DiPIT approach to decrease the {\color{red} memory consumption} of the PIT table using {\color{red} the Bloom} filter is proposed. The DiPIT implements small Bloom filters on each interface with a central Bloom filter to control false positives generated by the existing small Bloom filter on the interfaces.
 
In \cite{ref-new3}, an implementation of PIT using a new filter called FTDF-PIT is proposed. The FTDF-PIT uses a fast two-dimensional filter (FTDF) in {\color{red} the PIT} implementation. The FTDF performance is better than other proposed filters like Bloom and Quotient filters. The FTDF has higher performance in terms of insertion/deletion/ query in comparison to the Bloom and quotient filters. It also has lower memory consumption and a lower false positive rate when compared to the other mentioned filters. Therefore, these properties make it suitable for PIT in NDN networks.
In \cite{ref-new4}, the PIT table of a router is modeled using queuing theory to achieve {\color{red} the optimal} size. They evaluate the optimal PIT size to trade-off the network performance and {\color{red} the PIT table cost}. This model defines an objective function to minimize the PIT size subject to the interest drop probability as upper bound. 
In \cite{ref-new5}, an analytical model of two PIT replacement approaches is proposed. These approaches are called PIT replacement without reservation (PRWR) and PIT replacement with reservation (PRWR).
The first approach utilizes the PIT replacement, while the second approach uses the PIT entry reservation in conjunction with the first one. Both the approaches utilize  Continuous Time Markov Chain (CTMC) and improved the efficiency in comparison to the basic NDN.

{\color{red}
In \cite{ref-new7}, the 2DNCF-PIT to design the PIT table as a two-dimensional neighbor-based Cuckoo filter PIT table is introduced which improves the performance of PIT. In this work, the two-dimensional Cuckoo filter which uses the neighborhood buckets for the overflowed items is introduced. Using of 2DNCF in PIT design improves the performance of PIT in NDN networks.}

{\color{blue} In \cite{ref-new8}, another data structure based on the Cuckoo filter called CF-PIT is introduced. The CF-PIT is a two-dimensional Cuckoo filter that increases the size of the bucket in the standard Cuckoo filter using mini-buckets. The authors of this paper have claimed that the CF-PIT improves the performance of the Cuckoo filter. The CF-PIT has some main drawbacks as follows. The concept of slot and mini-bucket has not fully described. The authors mentioned that the slots number corresponds to the number of ports in the NDN-router, which each slot contains a mini-bucket, therefore, some issues is existed in this data structure, first, how the slot is selected, for example in a router with 16 ingress ports 16 slots need to be checked. The second issue is related to the structure of the mini-bucket. It seems that the size of the element of {\color{cyan} the mini-bucket} must be the same as the buckets of the standard Cuckoo filter, so the mini-buckets must be searched sequentially, these factors increase the lookup time. In addition, due to the existence of the slots and mini-buckets in the CF-PIT, the memory usage is higher than other proposed filters. The {\color{cyan}third} issue is that the paper did not mention the simulation parameters, for example, the mini-bucket size, the number of slots, the type of hashing functions, the finger print size, and the name of the dataset, therefore, simulating the CF-PIT based on the information in paper is not possible, due to the lack of {\color{cyan} the mentioned} parameters and finally, we could not compare our proposed work to the CF-PIT.  

 Another PIT design technique has been proposed in \cite{ref-new9} called MaPIT. MaPIT consists of two components: Index Table, and Packet Store. Index Table also has two data structures, a standard Bloom filter, and a mapping array. {\color{cyan} The Index Table is in on-chip memory, and the Packet Store} is in off-chip memory. Both the standard Bloom filter data structure and the mapping array are arrays of bits.
The mapping array has $n$ bits, and the standard Bloom filter is divided into $n$ parts. To add an item to this data structure, the item using the $k$ hashing function is first inserted into the Bloom filter. The number of bits that have been changed to bit one is entered in the mapping array. The number in the mapping array is where the item is stored in the Packet Store. This data structure supports deletion, but has additional calculations when searching and inserting items, and also has a high false positive rate.

In \cite{ref-new10}, dai et al. used a method called Name Component Coding (NCE). In this method, each name component of the Interest or Data packet is encoded by an NCE method, and then the generated code by the Encoding Name Prefix Trie (ENPT) method is converted to a Name Prefix Trie (NPT). This method has additional calculations, and the PIT table must also have a special architecture.

}
  Our proposed DiCuPIT approach utilizes the Cuckoo filters in each interface and {\color{red}the global} Cuckoo filter as a shared area, which achieves {\color{red} a lower} false positive rate and {\color{cyan} a higher performance} in comparison to other filters.

\section{Named Data Networking and Probabilistic Filters}
\label{NDN}
 In this section the concept of named data networking and probabilistic filters e.g. Bloom filter, Cuckoo filter is presented. 
\subsection{Named data networking}

\begin{figure}[h]
 \begin{small}
\begin{center}
\includegraphics[scale=0.23]{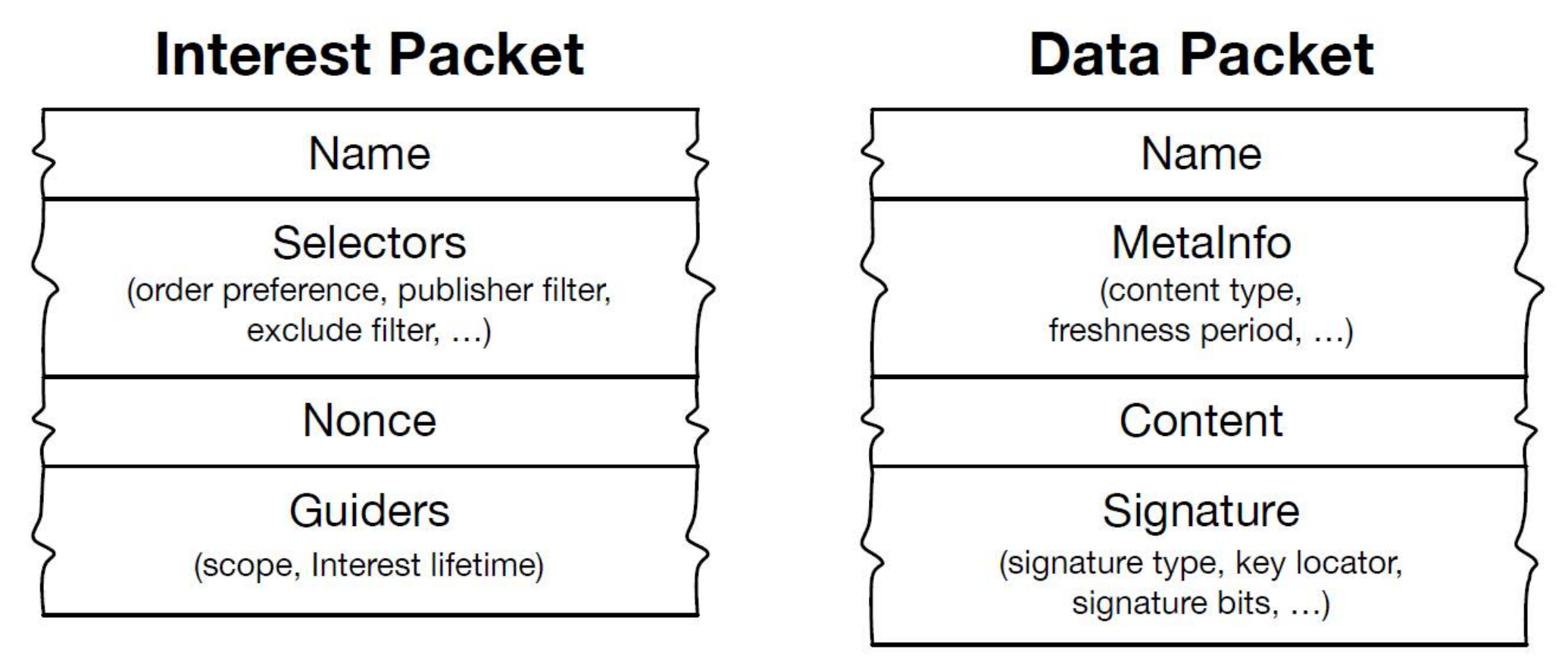}
\caption{The structure of NDN network packets \cite{ref4}.}
\label{fig:f00}
\end{center}
\end{small}
\end{figure}

In the NDN, the packets have a unique name {\color{red}and routing} is performed based on those packets \cite{ref-new9}. {\color{red} Names} in the named data networks are hierarchical, which will be separated by \enquote{/} characters based on the components \cite{ref1}. These names are like URLs that can be used to identify packets. For example, \enquote{/razi/ac/ir}  has three components, which are razi, ac and ir. This is an example of a name in data networks. The hierarchy of names in this architecture causes a lot of space to be wasted on memory. There are two types of packets in these networks: interest packets and data packets that both carry names and are identified based on these names. Figure \ref{fig:f00} depicts the structure of these packets.

When a data agent wants to create an Interest packet, {\color{cyan} puts} the name of the data in it and sends it to the network, and when the data producer receives an Interest packet, it creates a Data packet which includes names of the data and the required data. Puts it on the Data packet and sends it over the network. The network routing is done based on these names.

\begin{figure}[h]
 \begin{small}
\begin{center}
\includegraphics[scale=0.45]{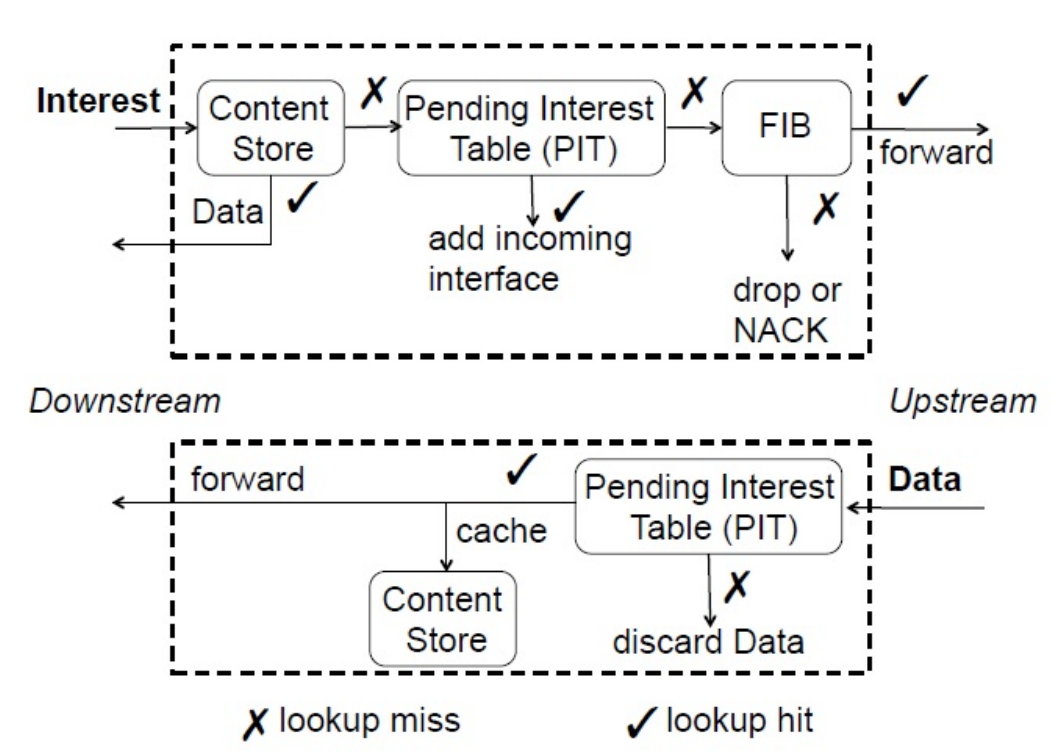}
\caption{NDN router strategy when receives an interest packet \cite{ref4}.}
\label{fig:f01}
\end{center}
\end{small}
\end{figure}

In NDN networking, each router contains three tables; Content Store (CS), Pending Interest Table (PIT), and Forwarding Information Base (FIB).  Figure \ref{fig:f01} depicts the architecture of an NDN router with its {\color{red}associated} tables and its strategy when receives an interest packet. 

In an NDN network, CS is the place of data packets that have already been processed in this router. PIT {\color{red} is supposed to maintain the} interest packets that have not yet received their response in the form of the data packet and finally FIB, which is used to forward the interest packets \cite{ref4}. The data consumer puts the intended content-name in an interest packet and sends it on the network. The NDN router first checks the CS. If it contains the data packet corresponding to the received interest packet, it will be forwarded directly to the consumer, which leads to an optimally use of network resources. Otherwise, the router searches the PIT; if it finds a name exactly the same as the content-name of the received interest packet in one of the entries, the PIT table will insert the incoming interface of the received interest packet inside the interfaces list and removes the packet. Otherwise, it will create a new entry for the received interest packet, then sends it to {\color{red} the FIB} for forwarding, if the router receives the data packet. At first, it will {\color{red}search} the PIT table in order to find the interest packet corresponding to the received data packet, then it forwards the data packet to its interfaces list, and removes the desired entry \cite{ref7}. 

\begin{figure}[h]
 \begin{small}
\begin{center}
\includegraphics[scale=0.75]{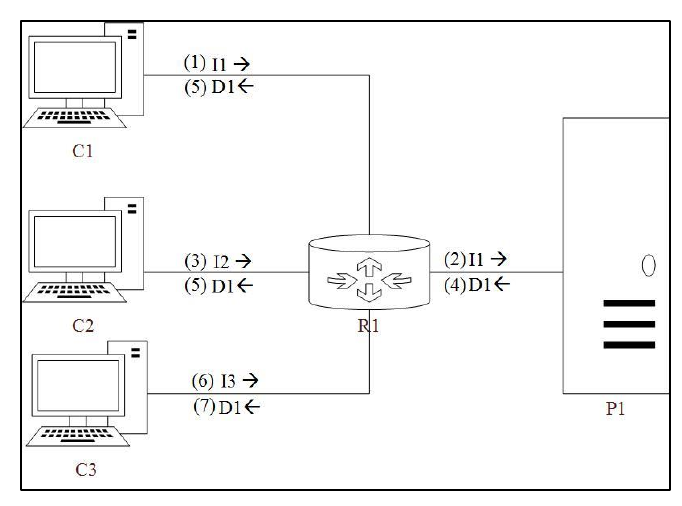}
\caption{NDN router strategy when receives an interest packet.}
\label{fig:f02}
\end{center}
\end{small}
\end{figure}

Eventually, it stores the received data packet in the CS table. {\color{red} The PIT} table must be able to perform the operations in a short time, support the timer, and have an algorithm to detect exactly the same \cite{ref7}. 
 Figure \ref{fig:f02} shows a scenario of a named data network. {\color{red}This example shows} three numbers of data consumers and {\color{red} multiple data producers named} P1, and also this scenario has a router {\color{red} named} R1. 

In step (1), C1 has a data request named \enquote{razi/ac/ir/eng/ computer-engineering.html}. C1 creates an Interest packet (I1) and sends the name of requested data through the network. When packet I1 reaches router R1, the router first checks its CS table, because it does not have packet data for I1, so it searches the PIT table and {\color{red} finds no} packet similar to I1, subsequently, it inserts the information of packet I1 into the PIT table and in step (2) delivers the packet to the FIB table for routing.  

In step (3), C2 also requests data called \enquote{razi/ac/ir/eng/ computer-engineering.html} and sends it as an Interest packet (I2) through the network. Router R1 first checks its CS table as soon as it receives the I2 packet because it does not find the data corresponding to the I2 packet, so it searches for the PIT table. Finds a packet similar to packet I2 in the PIT table. In this case, it {\color{red} simply} inserts the input interface number of packet I2 in information header of the packet I1.

In step (4) of router R1, {\color{red}the data packet (D1) receives} \enquote{ razi/ ac / ir / eng / computer-engineering.html}. First, to reuse the D1 packet, the packet is stored in CS and then searches the PIT table. Because there is an entry from the table that has the same name as D1, the router notices that there are Interest packets that requested D1 from this router. In this case, the router sends the D1 packet to the list of interfaces found in the entry, which in this scenario are interfaces 1 and 2. The entry in the PIT table is then deleted.

In step (6) C3 sends a request called \enquote{razi / ac / ir / eng / computer-engineering.html} in the form of the Interest packet (I3) to R1 router. In R1, upon receiving I3, it first searches the CS table because the Data packet (D1) corresponding to I3 is available in the CS table. In step (7), the router immediately sends the I3 packet in the opposite direction to C3.

\subsection{Probabilistic filters}
In this subsection, the probabilistic data structures which also called approximate membership checking data structures including {\color{red} the Bloom filter and the Cuckoo filter} is introduced.
\subsubsection{Bloom filter}
\begin{figure}[h]
 \begin{small}
\begin{center}
\includegraphics[scale=0.65]{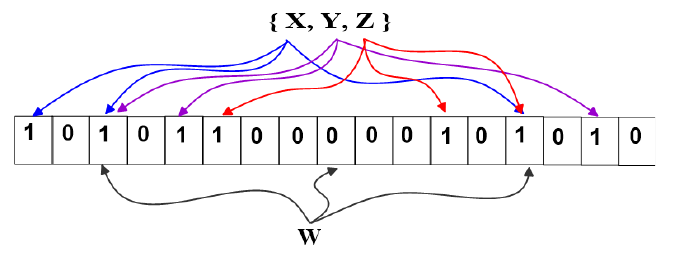}
\caption{A simple example to add three items $\{X,\ \ Y,\ \ Z\}$ to Bloom filter, and querying the Bloom filter for an item $W$.}
\label{fig:f03}
\end{center}
\end{small}
\end{figure}

{color{red} the Bloom} filter (BF) \cite{ref-new5} is the most popular probabilistic data structure, used {\color{red} for} membership query in a set. 
{\color{red} The Bloom} filter consists of an $m$ bits bit-array, initially all set to zero and $k$ hash functions $\{h_1, h_2,..., h_k\}$, each one in range $\{0...,m-1\}$. In a Bloom filter, $n$ items belong to a set $S$, where $S = \{x_1,x_2,...,x_n\}$. A Bloom filter is constructed in two phases: programming and {\color{red} querying}. In the programming phase, each item $x\in S$, $k$ hashing functions is calculated and the corresponding address in bit-array is set to one. In the {\color{red} querying} phase, for an item $z$, $k$ hashing functions {\color{red} are calculated} and the corresponding bits in {\color{red}the bit-array} are checked. If all the bits have the value one, then the item {\color{red}exists} in the bit-array with high probability, otherwise the item is not {\color{red} a member} of a set $S$. The false positive {color{red} can exist in a Bloom filter depending on the size of the bit-array $m$, the number of items $n$, and the number of hashing functions $k$.} The false positive probability is calculated as: $P\textsubscript{f}$=${(1-e^{-k*(n/m)})}^k$. Figure \ref{fig:f03} depicts a simple example {\color{red} for inserting three items $\{X,\ \ Y,\ \ X \} $ into BF}, and querying the BF for an item $W$.


\subsubsection{ Cuckoo filter }

The Cuckoo filter (CF) \cite{ref9} is a fast approximate membership query data structure {\color{red}in which} elements can be inserted and deleted dynamically in $O(1)$ time. The Cuckoo filter supports three main operations: insert, query, and delete. It builds on the main idea of the Cuckoo hash table, which consists of an array of buckets {\color{red}in which} two hash functions are used to generate the address of two candidate buckets for an element.

To lookup an element, both buckets are examined to check if either includes this element. To insert an element, both buckets are considered, and if any of them is empty, then the element is inserted into that vacant bucket. If both buckets already contain elements, the element chooses either of these buckets, moving out the already inserted element and re-inserts this element to its alternate position. 
This process {\color{red}an be repeated until} an unoccupied bucket is discovered, or {\color{red} until} reaching the upper limit of displacements. 
 For each element $x$, this hashing method computes the addresses of the two candidate buckets {\color{red}as follows}:
\begin{eqnarray}
f& =& fingerprint(x) \\
h_1(x)& =& hash(x) \\
h_2(x) &=& h_1(x) \oplus hash(f)
\label{Eq1}
\end{eqnarray}
Utilizing xor operation in Eq. \ref{Eq1} will guarantee a significant feature: $h1(x)$ can also be computed from $h2(x)$ and the fingerprint by employing the same equation.

\begin{figure}[h]
 \begin{small}
\begin{center}
\includegraphics[scale=0.7]{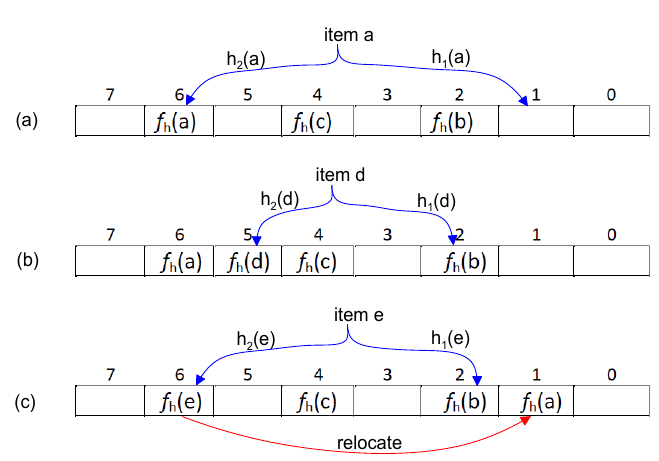}
\caption{(a) Inserting a at bucket (6), (b) Inserting d at bucket (5), (c) Inserting item e at
bucket(6).}
\label{fig:f04}
\end{center}
\end{small}
\end{figure}

A Cuckoo filter with 8 buckets is depicted in Figure \ref{fig:f04}. In this example, each bucket can store one item ($k=1$).

Initially, two items $b$, and $c$ have been inserted in the Cuckoo filter. {\color{red} The insertion} of the item \enquote{a} into Cuckoo filter is depicted in Figure \ref{fig:f04} (a). Therefore, the calculated addresses are: $h_1(a) = 1$, and $h_2(a) = 6$, and both buckets are checked, if one of them is empty then $f_h(a)$  is inserted into that empty bucket. In this case, both $h_1(a) = 1$, and $h_2(a) = 6$ buckets are empty, then one of them is selected randomly (bucket(6)) and $f_h(a)$  is inserted into it. Insertion of the item \enquote{d} into Cuckoo filter is depicted in Figure \ref{fig:f04} (b). Therefore, the calculated addresses are: $h_1(d) = 2$, and $h_2(d) = 5$. In this case, one bucket is empty (bucket(5)), then $f_h(d)$ is inserted into that empty bucket.
{\color{red} The insertion} of the item \enquote{e} into {\color{red} the Cuckoo} filter is depicted in Figure \ref{fig:f04} (c). Therefore, the calculated addresses are: $h_1(e) = 2$, and $h_2(d) = 6$. In this case, both buckets are occupied, then it is inserted into one of that buckets (bucket 6), kicks out the existing item (\enquote{a}), this old item is reinserted into its own alternate location (bucket 1). The process of ejecting of existing items and reinserting them in another locations called {\color{cyan} relocation} process.

 In some cases, {\color{cyan} relocating} the old item may also require ejecting another existing item, and this procedure may repeat until an unoccupied bucket is found or the maximum number of {\color{cyan}relocations} is reached (e.g., 150 times in our implementation). Even with long sequence of {\color{cyan}relocation} process, {\color{red} the Cuckoo} filter still has $O(1)$ insertion time and high space occupancy.


To lookup {\color{red} the Cuckoo} filter for a given item $x$, first calculate $x$'s fingerprint $f_h(x)$ and two candidate buckets $h_1(x)$, and $h_2(x)$, and then read these two buckets. If there is fingerprint match $f_h(x)$ in these two buckets then returns true. Otherwise, returns false.
The Cuckoo filter has {\color{red} other useful} features such as: it ensures that there is no false negatives as long as bucket overflow never occurs, it has better lookup performance and it supports deleting items dynamically \cite{ref-new7}.

\section{DiCuPIT: Distributed {\color{cyan}Cuckoo Filter-based} PIT}
\label{dicupit}

\begin{figure}[h]
 \begin{small}
\begin{center}
\includegraphics[scale=0.4]{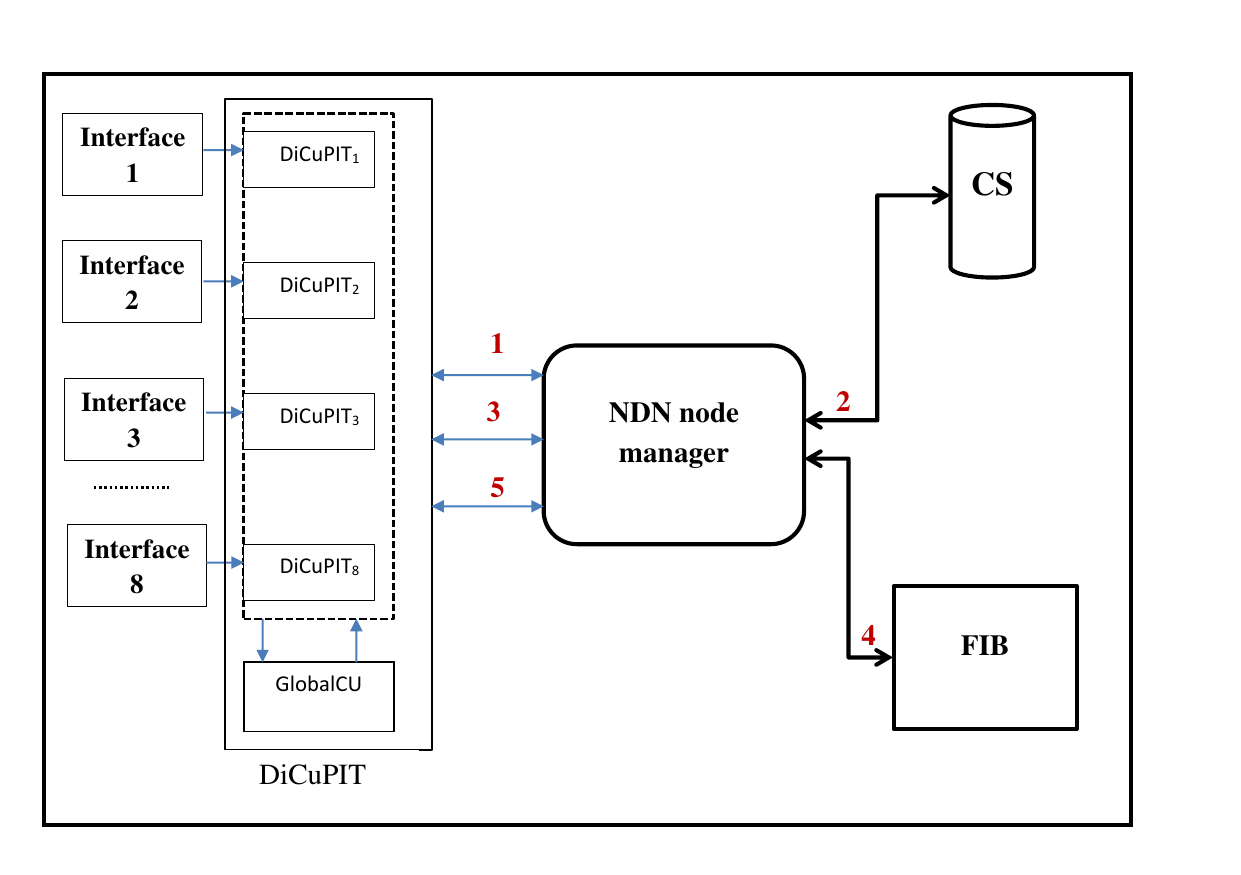}
\caption{A general view of the NDN router with eight interfaces, the proposed DiCuPIT table and different steps of interest packet arrival.}
\label{fig:f05}
\end{center}
\end{small}
\end{figure}

In this section, a new PIT table is proposed based on the Cuckoo filter, which can resolve the requirements of the PIT table according to the changes made. The Cuckoo filter is selected for this table because it has {\color{red} a better} lookup time compared to other filters such as the Bloom filter, and it can reduce response time in NDN networks. Another reason to use the Cuckoo filter is supporting of removing, which is important in the PIT table. DiCuPIT contains two types of sub-table; one of them for each interface that is called $DiCuPIT_i$ and another that is common between all available $DiCuPIT_i$s in DiCuPIT, which is called GlobalCu.
 Figure \ref{fig:f05} depicts a general view of an NDN router with eight {\color{red}interfaces} and the proposed DiCuPIT table inside it. It also shows the different steps {\color{red} involved in the} arrival of an interest packet.

In Figure \ref{fig:f05}, {\color{red} in the first step}, an interest packet enters from an interface, in the second step, it is checked against the content store, if a match is found the requested data packet is fetched and{\color{red} the NDN} node responds by sending back the data packet through the interface which the interest packet has received. Otherwise, in the third step NDN name is added to the DiCuPIT table and in the fourth step {\color{red} the Interest} packet {\color{cyan} will be sent} to the FIB table, and in the last step it is forwarded to its destination

\subsection { {\color{cyan}Structure of interface's sub-tables }}

Each $DiCuPIT_i$ is a table containing $e$ buckets, and each bucket contains $b$ entries. Each entry in $DiCuPIT_i$ includes the fingerprint and expiration field and does not require storing the interface-list, since a $DiCuPIT_i$ is considered for each interface in the router. Each $DiCuPIT_i$ sub-table works based on the Cuckoo filter, using the name of the incoming interest packet, the memory position of the packet in the table, and {\color{red} the fingerprint} will be computed. If it is not in the table, it will be stored in the fingerprint field and its expiration date in the expiration field. For {\color{red} the incoming} data packet is also computed by the name of memory position and {\color{red} the fingerprint}. If there is a similar fingerprint in calculated position, the incoming data packet will be forwarded from the related face. All of $DiCuPIT_i$s are the same and using a kind of hash function, but they work independently. This approach helps to make a general lookup among all $DiCuPIT_i$s. In this way, by doing hash on $DiCuPIT_i$ it is possible to {\color{red} know} the existence of the name of the same packet in another $DiCuPIT_i$. This is done {\color{red} using} the GlobalSearch algorithm. 

\subsection{GlobalCu {\color{cyan}subtable} structure}
\begin{figure}[h]
 \begin{small}
\begin{center}
\includegraphics[scale=0.9]{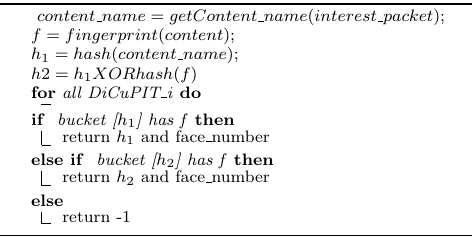}
\caption{GlobalSearch (packet).}
\label{algo1}
\end{center}
\end{small}
\end{figure}

{\color{red}

{\color{cyan}In DiCuPIT, a table is used for the connection between $DiCuPIT_i$s in PIT which is called GlobalCu.} The GlobalCU works based on {\color{red} the Cuckoo} filter. The GlobalCU and $DiCuPIT_i$ have {\color{red} the same size}, this means that GlobalCU includes $e$ buckets, each bucket has the $b$ entry. The only difference between GlobalCU and $DiCuPIT_i$ is that the GlobalCU stores the list of incoming interfaces. When an interest packet enters the named data network router, the received information of the packet is stored in the corresponding $DiCuPIT_i$. The problem is that an interest packet with the same content name may have entered the router from another interface, {\color{cyan}in this case, the information} about the same interface is stored in two or more DiCuPITs, which consumes memory in the PIT table. To address this, DiCuPIT uses a common table between all DiCuPITs called GlobalCU. GlobalCu is the storage of interest packets {\color{cyan}coming into the router} from several different faces and is exactly the same as the sub-table, except that it also stores the list of interfaces. The GlobalCu sub-table in DiCuPIT supports the aggregation feature. {\color{cyan}This reduces the search time in this table but requires more memory.} The main idea of GlobalCu is that GlobalSearch can be done on all $DiCuPIT_i$ using Cuckoo filter features. This is because $DiCuPIT_i$ are exactly the same and use the same hash functions, so it can be said that when searching in a $DiCuPIT_i$, if the name is in the entry with the value $h$, it is existed in all $DiCuPIT_i$ in the entry in the number $h$, so in total the hash function only once, but it searches the whole $DiCuPIT_i$ and this is done in parallel. This algorithm receives an interest packet as input and calculates the fingerprint values $h1$ and $h2$, which represent the selected locations, because all sub-tables are the same. In this algorithm, if the desired value is found, then the value of $h1$ and the interface number are returned, otherwise the value of -1 is returned as {\color{cyan} absent}.
 The GlobalSearch pseudo-code} is shown in Fig. \ref{algo1}.

\subsection{Architecture of DiCuPIT }

\begin{figure*}
 \begin{small}
\begin{center}
\includegraphics[scale=0.4]{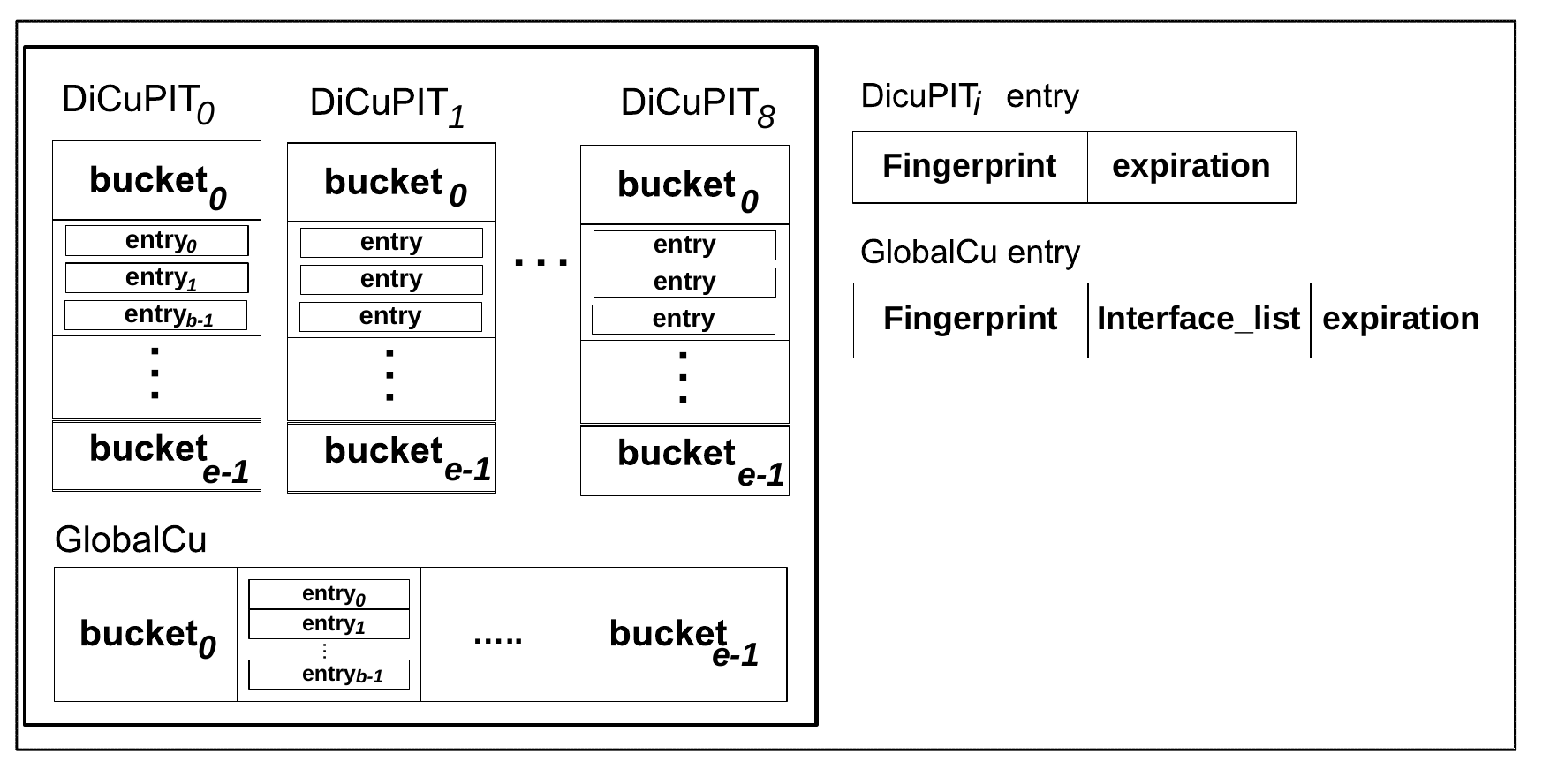}
\caption{DiCuPIT architecture for a NDN router with eight ports and fields of each entry.}
\label{fig:f1}
\end{center}
\end{small}
\end{figure*}

DiCuPIT is a distributed data structure that allocates the $DiCuPIT_i$ sub-table to each entry face of router. Each of $DiCuPIT_i$s stores the information related to the corresponding interface. For instance, $DiCuPIT_m$ stores the information of face number $m$. Given that DiCuPIT is distributed, so the memory consumption reduces in PIT because it is no longer necessary to keep the list of interfaces in the PIT table. DiCuPIT is a fast data structure with a low memory consumption that matches the PIT table requirement. In DiCuPIT designing the cases of integration, scalability and algorithms are considered to detect exactly the same. In original PIT, the information related to interfaces are centralized, which can guarantee the integration of PIT, but a table is considered for each face in DiCuPIT to reduce the redundancy of information related to interfaces-list which result in the reduction of memory consumption. As shown in Fig. \ref{fig:f1}, this schematic is intended for a router with eight ports which includes 8 $DiCuPIT_i$sub-tables and a sub-table named GlobalCu. In PITs containing distributed tables, the integration feature is not supported. To solve this problem. In DiCuPIT, another table named GlobalCu is used, which can guarantee that DiCuPIt supports the integration feature by using this table which makes lookup faster in PIT. 

\begin{figure}[h]
 \begin{small}
\begin{center}
\includegraphics[scale=0.9]{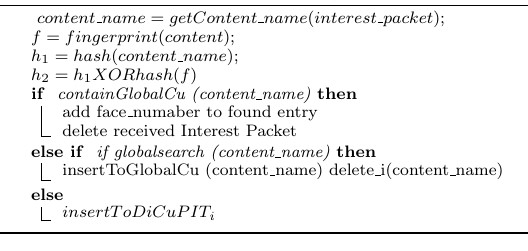}
\caption{ Received Interest Packet (interest\_packet).}
\label{algo2}
\end{center}
\end{small}
\end{figure}

\begin{figure}[h]
 \begin{small}
\begin{center}
\includegraphics[scale=0.9]{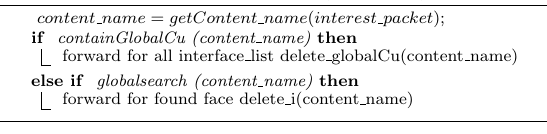}
\caption{Received Interest Packet (data\_packet).}
\label{algo3}
\end{center}
\end{small}
\end{figure}

When an interest packet enters the NDN router, the first lookup will be done on the GlobalCu table. If the same name as {\color{red} received interest packet} is found in this table, the interface information of this packet will be inserted in the GlobalCu table and the packet will be ignored; otherwise, a GlobalSearch is done on all $DiCuPIT_i$s to see whether the interest packet with the same name has been already entered {\color{red} the router} from another interface. If the answer is yes, the information related to the received interest packet will be stored in the GlobalCu table along with the numbers of interfaces which the packet has been entered from it. If the answer is negative, it means the received interest packet enters this router for the first time and {\color{red} the information} will be stored in dedicated $DiCuPIT_i$ for forwarding, the packet is delivered to the FIB table. Fig. \ref{algo2} shows the reaction of DiCuPIT when it receives an interest packet. 
 First, the packet name is received. The containGlobalCu function gets the name of the packet, then searches for it in GlobalCu. If it was in GlobalCu, {\color{red} the related} information to the field would be updated, and the received packet will be removed; otherwise, the entire $DiCuPIT_i$ sub-tables will be searched by the GlobalSearch technique. If found in some value, the received interest packet will be inserted in GlobalCu table along with the received interest and interfaces-list; otherwise, the received interest packet will be inserted in the corresponding sub-table to entry face. Fig. \ref{algo3} shows the reaction of DiCuPIT when a data packet is entered.

$~content\_name=getContent\_name (interest\_packet)$\;



 If the data packet enters the router, first, the GlobalCu table will be examined. In this case, it means the corresponding interest packet to received data packet has been entered {\color{red} the router} from several different faces and the data packet will be sent to {\color{red} the interfaces-list}. The intended entry will be removed and the data packet will be stored in {\color{red} the CS table.} If the intended name is not found in GlobaCu, it is possible to find the interface which receives the corresponding interest packet to the data packet, then it sends the packet, and removes the intended entry and the received data packet will be stored in CS. 

 Similar to the DiPIT \cite{ref5}, the DiCuPIT, utilizes a $DiCuPIT_i$ in each interface in conjunction with a GlobalCU as a shared Cuckoo filter.
In the DiCuPIT, for all tables including all $DiCuPIT_i$ and GlobalCU, only {\color{red} two hashing functions are} calculated. Therefore, the time complexity is $O(1)$. 

{\color{cyan} The lookup time complexity of DiCuPIT is O(1). This table is exactly like the DiCuPIT sub-tables, and once the functions $h1$ and $h2$ are calculated.}
{\color{red} By calculating the functions $h1$ and $h2$, the desired sub-table and the desired bucket can be accessed. For example, if the router has 8 sub-tables, the first 3 bits determines {\color{cyan}the sub-table, and} the remaining bits determines the desired bucket). With these descriptions, we can say that the search time is equal to $O(1)$. It should be noted each interface of the NDN router includes a DiCuPIT data structure that have been implemented in hardware and concurrently performs {\color{cyan} a lookup operation}. In the other words, the lookup time has $O(1)$ complexity while $k$ access is performed simultaneously.}

The number of accesses is $k$ access and one access for GlobalCU, therefore the order of number of accesses is $O(k+1)$ which is showed as $O(k)$, where $k$ is the number of interfaces of the NDN router. In DiCuPIT, all $DiCuPIT_i$ and GlobalCU utilize {\color{cyan} the same} hashing functions (two hashing functions). Therefore, the computational cost is lower than the DiPIT that uses 5 hashing functions for each Bloom filter. In addition, the Cuckoo filter used in the DiCuPIT make possible the removing items with lower false positive, while the Bloom filter used in DiPIT the removing item is not possible and has {\color{cyan} a higher} false positive rate in the {\color{cyan}same conditions as the} Cuckoo filter.
{\color{red} In addition, the DiPIT uses {\color{red} a Bloom filter} which each DiPIT utilized 5 hashing {\color{cyan}functions} for each Bloom filter {\color{cyan} which means the} time complexity is $O(5)=O(1)$. Furthermore, in an NDN router with $k$ interface based on the DiPIT, the number of accesses is $O(5k)=O(k)$. As expected for the analysis, the achieved results show that the DiCuPIT improves the lookup time 36\% in comparison to the DiPIT. }

\section{Evaluation}
\label{results}
To evaluate DiCuPIT, it will be compared to three techniques named chain hash in \cite{ref-new9}, a 32-bit hash for centralized PIT, and DiPIT in \cite{ref5}.
To implement DiCuPIT, an NDN network is considered with three routers, 8 ports, four nodes, and the incoming packet rate for each port are assumed between 10Mpcs (Mega package per second) to 100Mpcs.  It should be noted that the network topology is useful when the routing algorithms need to be evaluated. In this case which a table like PIT is evaluated a huge number of names is needed. Therefore, we used a simple network topology with three routers.

 The size of $DiCuPIT_i$ contains the numbers {\color{red} from 8000 to 8000000 entries based on the incoming bandwidth.} Finally, the intended RTT is 80ms in these experiments \cite{ref10}. The used dataset for the experiment of the CNC dataset is the one used in \cite{ref11}. 
The specifications of the system used in this work {\color{red} are}:  AsusX299 Delux h1 equipped with Core i-9 7900 CPU, 64 GB DDR4 3200 g-skill (4*16)RAM, and 1080 11GB DDR5 GPU. The Python programming language has been used in the implementation. The bucket and fingerprint sizes have been set to 4 entries and 6 bits\cite{ref9}, respectively.

{\color{red} The Table} \ref{table1} shows the simulation parameters and {\color{red} the system specification} which used in this work.

\begin{table*}
\centering
\begin{small}
\caption{Simulation parameters.}
  \label{table1}
\begin{tabular}{|c|c|} \hline
Number of routers & 3 \\ \hline
Number of ports in each router& 16 \\ \hline
Packet rate of each port& 10MpcS-100 Mpcs \\ \hline
Number of entries of DiCuPIT & 8000-8000000 \\ \hline
Round trip time (RTT) & 80 Msec \\ \hline
Dataset & CNC dataset \cite{ref11} \\ \hline

\hline
 system specification& AsusX299 delux h1 \\ 
  & equipped with Core i-9 7900 CPU \\
  &64 GB DDR4 3200 g-skill (4*16)RAM \\
   & 1080 11GB DDR5 GPU  \\ \hline
 Programming language&   Python\\ \hline
Bucket size  &  4 entries \cite{ref9} \\ \hline
 Fingerprint size & 6 bits \cite{ref9}  \\ \hline
 Expiration size & 16 bits \\ \hline
Hash function of chain hashing  &MurMur (32 bits) \\ \hline
Number of hashing function in DiPIT & 5 \\ \hline
 
\end{tabular}
\end{small}

\end{table*}

\subsection{Evaluation of memory consumption}
\begin{figure}[h!]
 \begin{small}
\begin{center}
\includegraphics[scale=0.5]{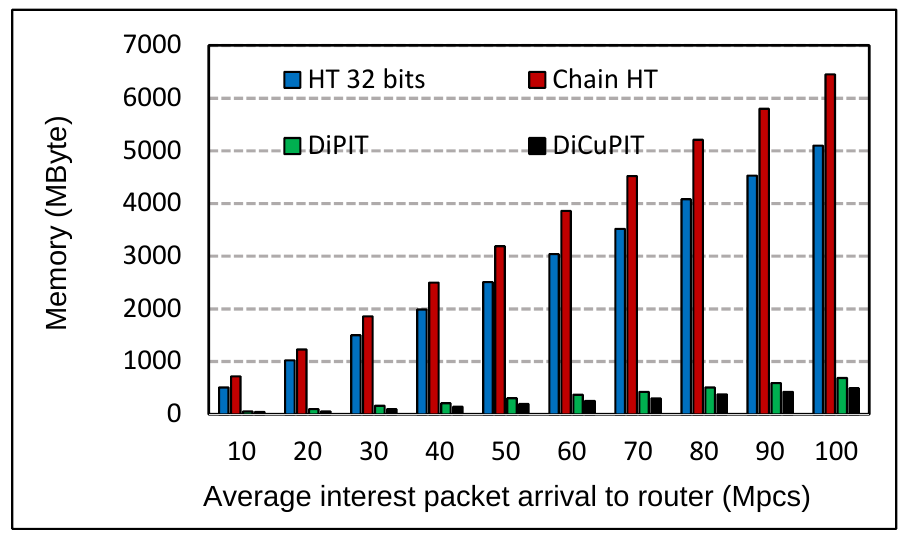}
\vspace{-0.1cm}
\caption{Memory used by DiCuPIT and its comparison with other methods.}
\label{fig2}
\end{center}
\end{small}
\end{figure}

To calculate the memory consumption. First, the incoming rate of different interest packets is considered for each interface.
Generally, for a centralized PIT, the amount of required memory without considering the CS is equivalent to $RTT*bandwidth$ based on the thumb rule \cite{ref5}, which RTT is round tripe time.
While the list of interfaces {\color{red} is} not stored in $DiCuPIT_i$, however the values of 6-bit as fingerprint and 16-bit for the expiration in $DiCuPIT_i$ are considered for each entry. In GlobalCu 6-bit fingerprint, 16-bit for expiration, and 8-bit for interfaces list are assumed. In doing the experiments a 32-bit hash function and 16-bit for expiration, 8-bit for {\color{red} the interfaces list} of each entry and finally another 32-bit hash for chain is used. For HT32, a 32-bit hash with 16-bit expiration and 8-bit for interfaces list is used as well. Also, for DiPIT, a Bloom filter with seven hash functions for each $PIT_i$ is used. In the chain hash method, to make the chain, the value of another hash is also stored, which shows the next value of the other chain, and more memory is consumed. As shown in Fig. \ref{fig2}, DiCuPIT consumes less memory.


Table \ref{table2} depicts memory consumption of DiCuPIT in compared to DiPIT as one the most relevant works and its improvement rate.

\begin{table}
\centering
\begin{small}
\caption{Memory consumption of DiCuPIT in compared to DiPIT and its improvement rate.}
  \label{table2}
\begin{tabular}{|p{2cm}|p{2cm}|p{2cm}|p{2cm}|} \hline

 Packet arrival rate (Mpsec)&	DiPIT memory consumption	& DiCuPIT memory consumption	& Memory improvement rate (\%) \\ \hline
10 &	55	&45	&18 \\ \hline
 20	&98	&56&	42\\ \hline
30&	164&	96&	41\\ \hline
40&	210&	145&	30\\ \hline
50	&305&	197	&35\\ \hline
60	&372&	256&	31\\ \hline
70	&426	&300&	30\\ \hline
80&	510&	380&	25\\ \hline
90&	593&	426	&28\\ \hline
100&	689&	495	&28\\ \hline

\end{tabular}
\end{small}

\end{table}

As can be seen in Table \ref{table2}, the average lookup time improvement of DiCuPIT in comparison to DiPIT is 31\%. This is because, the DiPIT utilizes five hashing functions, therefore, it needs five accesses to the bit-array of Bloom filter and needs another hashing function to access the packet in hash table, while the DiCuPIT utilizes at most two hashing functions which result two accesses.

\subsection{Lookup time} 

\begin{figure}[h]
 \begin{small}
\begin{center}
\includegraphics[scale=0.6]{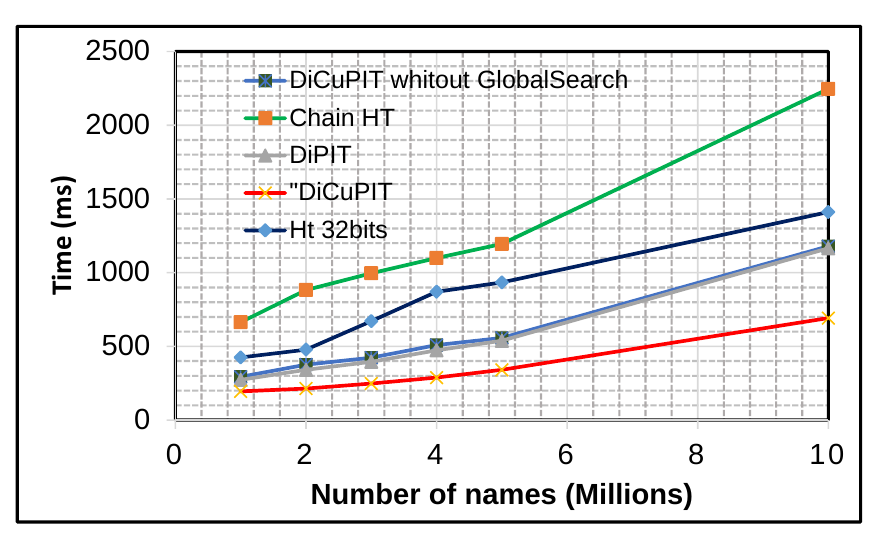}
\vspace{-0.1cm}
\caption{Average lookup time in DicuPIT and other methods.}
\label{fig3}
\end{center}
\end{small}
\end{figure}

{\color{red} The lookup} time is expected to decrease in DiCuPIT. At first, {\color{red} the interest packets} are inserted in PIT tables which are implemented with different techniques and a lookup is done on them. Fig. \ref{fig3} shows that the DiCuPIT method with the GlobalSearch technique has the least time in the chain hash experiment, because the hash function should be applied twice on names and the created chain contains the most lookup time. HT32bit-technique has a high lookup time as well because it probably searches another entry for searching elements. In DiPIT, seven hash functions are used for the elements. DiPIT has better lookup time compared to DiCuPIT without the GlobalSearch technique, and eventually, DiCuPIT with the GlobalSearch technique explained above, has better time compared to other mentioned techniques.

Table \ref{table3} depicts lookup time of DiCuPIT in compared to DiPIT as one the most relevant works and its improvement rate.

\begin{table}
\centering
\begin{small}
\caption{Lookup time of DiCuPIT in compared to DiPIT and its improvement rate.}
  \label{table3}
\begin{tabular}{|p{2cm}|p{2cm}|p{2cm}|p{2cm}|}  \hline

 Number of names (Million)&	DIPIT	& DiCuPIT& Improvement rate(\%)	 \\ \hline
1	&  272.8614  &	194.6903  &	28  \\ \hline
2   &	340.708	&213.8643&	37  \\ \hline
3&	 395.2802 &	247.7876 &	37 \\ \hline
4&	473.4513&	288.2006&	39 \\ \hline
5&	538.3481	&340.708	&37 \\ \hline
10&	1165.192	&691.7404&	40 \\ \hline

\end{tabular}
\end{small}
\end{table}

As can be seen in Table \ref{table3}, the average lookup time improvement of DiCuPII in comparison to DiPIT is 36\%. This is because, the number of hashing functions in DiPIT is five while the number of hashing functions in the DiCuPIT in the worst case is two.
From {\color{red} the Tables \ref{table2}, and {table3},} it can be observed that the DiCuPIt can improve both memory consumption and {\color{red} the lookup time} in comparison to the DiPIt approach. In addition, the improvement rate of lookup time of the DiCuPIt is higher than the memory consumption rate.

\subsection{ False positive evaluation}

\begin{figure}[h]
 \begin{small}
\begin{center}
\includegraphics[scale=0.45]{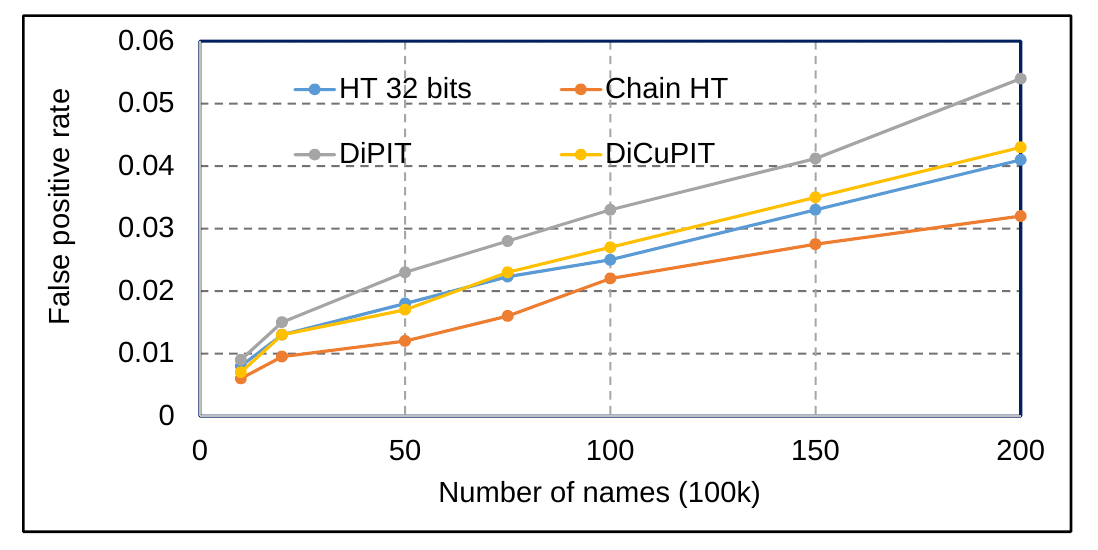}

\caption{ Possible false positives in DiCuPIT and other methods.}
\label{fig:f4}
\end{center}
\end{small}
\end{figure}

Two types of false positive may occur in NDN networks \cite{ref5}.
When an interest packet enters the router, if it finds a similar incoming packet in one face in the table by mistake, the incoming interest packet will not be forwarded, and the packet will be lost. This type of false positive is critical.
If the data packet enters the router, the router finds the faces at first, which the corresponding interest packet to this data packet enters the router from them, then forwarded the data packet to these faces. If the interest packet corresponding to incoming data packet is found in one face by mistake, this data packet will be sent in a fake way and the related information to the interest packet will be removed, so when the real data packet reaches the router, it does not find the interest packet and the incoming data packet will not be forwarded from the face. This false positive type is not critical because the extra propagate has no harm, just using network resources a little more. So, only the loss of the interest packet has been investigated in DiCuPIT. As seen in Fig. \ref{fig:f4} for two million records, it contains less than 1\% false positive, which is acceptable for a filter. For this, PIT tables are implemented at first with different techniques, and different names are inserted in the table, at the end, to obtain false positives, the query is taken from PIT tables. 
Since the chain hash resolves the collision in the form of a chain contains the least false positive, DiCuPIT also has an acceptable value of less than 0.1\%. 

{\color{red} From the evaluation results, the DiCuPIT improves the false positive rate \%19 in comparison to DiPIT. It should be noted that the precise theoretical analysis for DiCUPIT is not easy, but as it can be observed, the DiPIT uses a counting Bloom filter {\color{cyan}whose false positive result is identical to the standard Bloom filter, } while the DiCuPIT uses a set of the Cuckoo filters which has the lower false positive than the Bloom filter and finally the DiPIT. This means that the false positive of the DiCuPIT is lower than the DiPiT in the {\color{cyan} analytical} aspect too. }

\section{Conclusion}
\label{conclusion}
In this paper, a new data structure named DiCuPIT for the PIT table was presented in NDN networks which could reduce lookup time and memory consumption. In addition, the false positive value is acceptable in this data structure for the named data network. In DiCuPIT, the distributed tables are used for each face, and another sub-table called GlobalCu is used to guarantee the integration. These sub-tables and {\color{cyan} the GlobalCu} sub-table are the same. The risk in DiCuPIT is due to the relation of the GlobalCu sub-table with other sub-tables, which may be filled in before the other sub-tables. This risk can be overcome if the size of the sub-tables be assumed large enough. On the other hand, the sub-tables can be scaled and set to a larger size, gradually, for the possible overflow of the GlobalCu sub-table.


\nocite{*} 

\bibliographystyle{unsrtnat}
\bibliography{reference}

%

\end{document}